\documentclass[%
 reprint,
superscriptaddress,
 amsmath,amssymb,
 aps,
]{revtex4-2}

\usepackage{graphicx}
\usepackage{dcolumn}
\usepackage{bm}

\begin{document}

\preprint{APS/123-QED}

\title{Homophilic organization of egocentric communities in ICT services}

\author{Chandreyee Roy}
\affiliation{
Department of Computer Science, Aalto University School of Science, Espoo, Finland
}%

\author{Hang-Hyun Jo}%
\email{h2jo@catholic.ac.kr}
\affiliation{%
Department of Physics, The Catholic University of Korea, Bucheon, Republic of Korea
}%

\author{J\'anos Kert\'esz}
\affiliation{%
Department of Network and Data Science, Central European University, Vienna, Austria
}

\author{Kimmo Kaski}
\affiliation{
Department of Computer Science, Aalto University School of Science, Espoo, Finland
}%

\author{J\'anos T\"or\"ok}
\affiliation{
Department of Theoretical Physics, Budapest University of Technology and Economics, Budapest, Hungary
}
\affiliation{
Institute of Physics, Budapest University of Technology and Economics, Budapest, Hungary
}
\affiliation{
MTA-BME Morphodynamics Research Group, Budapest University of Technology and Economics, Budapest, Hungary
}
  



\date{\today}

\begin{abstract}
Members of a society can be characterized by a large number of features, such as gender, age, ethnicity, religion, social status, and shared activities. One of the main tie-forming factors between individuals in human societies is homophily, the tendency of being attracted to similar others. Homophily has been mainly studied with focus on one of the features and little is known about the roles of similarities of different origins in the formation of communities. To close this gap, we analyze three datasets from Information and Communications Technology (ICT) services, namely, two online social networks and a network deduced from mobile phone calls, in all of which metadata about individual features are available. We identify communities within egocentric networks and surprisingly find that the larger the community is, the more overlap is found between features of its members and the ego. We interpret this finding in terms of the effort needed to manage the communities; the larger diversity requires more effort such that to maintain a large diverse group may exceed the capacity of the members. As the ego reaches out to her alters on an ICT service, we observe that the first alter in each community tends to have a higher feature overlap with the ego than the rest. Moreover the feature overlap of the ego with all her alters displays a non-monotonic behaviors as a function of the ego's degree. We propose a simple mechanism of how people add links in their egocentric networks of alters that reproduces all the empirical observations and shows the reason behind non-monotonic tendency of the egocentric feature overlap as a function of the ego's degree.
\end{abstract}

\maketitle

\section{Introduction}

One of the central questions in sociology is how the social network of a person is organized. The ties are known to stem from the kinship, triadic closure, and in modern times to a large extent from  homophily, i.e., the tendency of similar individuals getting associated with each other~\cite{McPherson2001Birds}. Such association is related to single or multiple features of individuals like the race, gender, age, location, level of education, social or economic status, hobby, etc. In short whatever social activity we do we meet others with similar social preference, and with whom we can make social connections.

An egocentric network consists of an ego and her alters. Alters are expected to share some of the features with the ego. Such an egocentric network is structured, namely, its members form communities corresponding to different types of groups the ego belongs to. Through the ego this produces an overlapping community structure~\cite{Palla2005Uncovering, Ahn2010Link}. The present study focuses on investigating the interplay between the heterogeneity of links and community structure in egocentric networks. Robin Dunbar has proposed other organisational structures within the network of alters, namely the Dunbar's circles, and argues that humans have on average $150$ active social contacts~\cite{Dunbar1995Social, Dunbar2011Constraints}. Other acquaintances we connect to at Online Social Network (OSN) sites are people we know but have no social relation. On the other hand the $150$ alters are not equally important to us; there are $\sim$$5$ loved ones, $\sim$$15$ close friends, and $\sim$$50$ friends. In addition, individuals in a network belong to multiple social groups simultaneously and tend to connect with each other based on several sociological aspects such as gender~\cite{Kovanen2013Temporal,laniado2016gender}, ethnicity~\cite{shrum1988friendship,wimmer2010racial}, shared activities~\cite{feld1982social} among others. Having multiple attributes in common does not necessarily mean that a pair will form a strong bond when compared to cases where only one attribute is common~\cite{block2014multidimensional}. Traditional research on homophily has mostly been concentrated on one dimension~\cite{Kovanen2013Temporal, feld1982social, stehle2013gender} but considering the different attributes of individuals can capture more complex friendship mechanisms~\cite{hooijsma2020multidimensional, lorenz2020social}. For example, a previous study on school networks in Europe has found that the preference in friendship is not just based on ethnicity but is dependent on gender, socioeconomic status, academic achievements, and cultural traits among others~\cite{campigotto2022school}. Furthermore, another study in the USA has revealed that increasing racial diversity in schools does not necessarily lead to more inter-racial friendships but it needs to include several different factors such as school environment, status equality, and extracurricular activities to boost less segregation in friendship choices~\cite{moody2001race}. Even behavioral patterns within social media platforms require a multidimensional approach to understanding homophily~\cite{nahon2014homophily}. 

Nowadays, in order to keep up our social ties we use various Information and Communications Technology (ICT) services, through which we can connect to our friends. The order in which they are reached tells a lot about their importance to us. This could be related to the emotional closeness~\cite{Dunbar2003Social} or the community structure~\cite{Granovetter1973Strength}, or both of them. This is a question of importance that has not yet a clear answer.

In our earlier studies~\cite{Torok2016What, Murase2019Sampling} we showed that the observed network properties could be strongly influenced by the inherent sampling process stemming from the usage of data from single communication channels. Therefore, this type of sampling might be closely related to the way people choose their channels of communication. For example, users of ICT facilities have in general a variety of different services from which they can choose. As a result of such choices not all social links show up in each ICT service. Thus egocentric networks of the users in an ICT service channel must be partial.

\begin{table*}[t!]
\centering
\caption{Metadata of the available datasets. Features are mostly self-declared and their availability is shown in percentage of all users in each dataset. The location refers to city for iWiW and one of the 188 Slovakian regions for Pokec. Features marked with `cat' are categorical variables and those marked with `num' are numeric variables.}
\begin{tabular}{|l||r|c|r|c|r|c|}
	\hline
	Feature & \multicolumn{2}{c|}{iWiW}  & \multicolumn{2}{c|}{Pokec}  & \multicolumn{2}{c|}{CDR} \\
	\hline
Age	& 61\% & num  & 61\% & num & 86.5\% & num  \\
	\hline
Gender	& 100\%  & cat & 100\% & cat & 86.5\% &cat\\
	\hline
Location	& 89\%  & cat & 100\%& cat & 
& 
\\
	\hline
Education level	& 53\%& cat && &  &  \\
	\hline
Body Mass Index (BMI) &  & & 40\% & num & & \\
	\hline
Alcohol consumption &  & & 48\% & cat &  &\\
	\hline
Sexual preference	&   && 31\% & cat & & \\
	\hline
Smoking 	& &  & 54\% & cat &&  \\
	\hline
\end{tabular}
\label{table:1}
\end{table*}

The channel selection is a complex decision making process that has an individual and a social component. When modeling this process to understand the consequences of the implied sampling bias, the assumption was employed in Ref.~\cite{Torok2016What} that the friends choose the service, which is the least inconvenient for both of them. However, this does not answer the temporal aspect of the question, namely, that in what order we reach our friends in a new service. Is it on average in the order of emotional closeness or do people have other preferences? The emotional closeness~\cite{Roberts2011Communication} is a complicated concept and measuring it in datasets with no link weight information or self-reported value is tricky~\cite{merritt2005investigation}. For this, the similarity of social features can be used as a proxy for the emotional closeness~\cite{reagans2011close}, which in turn can be measured if social traits are available.

This paper is organized such that we first discuss three ICT datasets to be analyzed. After introducing the measures for the analysis, we present our empirical findings for one of three datasets, followed by a model study for the comparison with the analysis results. Finally, we conclude our work with discussions and present the results of two other datasets for comparison purposes. 

\section{ICT data}

The datasets of relational databases we have access to and study are (i) iWiW, a Hungarian Online Social Network (OSN)~\cite{Lengyel2015Geographies}, (ii) Pokec, a Slovakian OSN~\cite{Takac2012Data}, and (iii) mobile phone data from a European country~\cite{onnela2007structure}. The first two databases contain user-acknowledged friendship links, while the third one is created by aggregating Call Detail Records (CDRs). The iWiW was once the most popular OSN service for three years in Hungary; it had 4.6 million users, and an average degree was $208$. The Pokec had similar success in Slovakia with 1.6 million users and an average degree of $27.3$. The mobile phone data is from a provider in a European country with $5.7$ million users. All three anonymized datasets are accompanied with metadata, summarized in Table~\ref{table:1}.

\begin{figure*}[!t]
    \centering
    \includegraphics[width=0.8\textwidth]{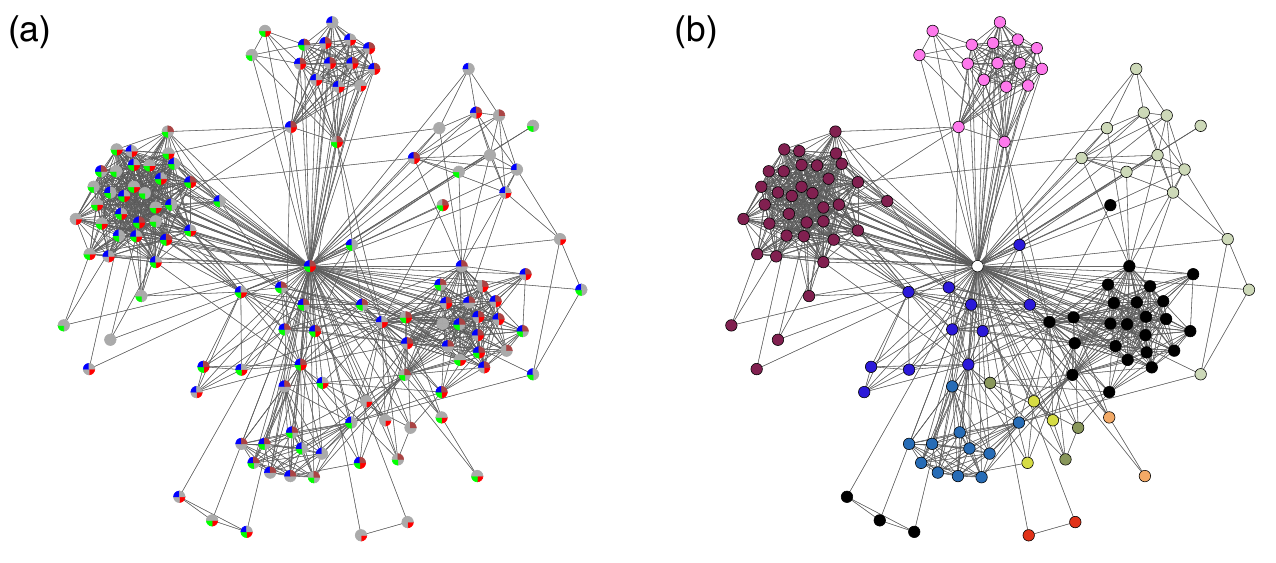}
    \caption{An exemplary egocentric network from the iWiW data. Links between the ego (central node) and her alters as well as between the alters are shown. (a) Four sectors in each circle represent four features available in the data. Whenever an alter has the same trait as the ego for a feature, that sector of the alter is colored as the ego (gender in red, city in green, education level in blue, and age in brown). Otherwise, if the alter's trait does not match with that of the ego, it is colored gray. (b) The same egocentric network in (a) is shown with communities detected by Infomap method~\cite{rosvall2009map}, where different communities are denoted by different node colors.}
    \label{fig:egomy}
\end{figure*}

The metadata of each dataset contains several features of users, and each feature may have different values or traits. For example, there are two traits, ``male" and ``female", allowed for the feature ``gender". A feature is considered overlapping for a pair of connected users if those users' traits for that feature match each other. For some features having continuous values of traits, we allow for value differences, e.g., we consider that two users are of the same age if the difference in their birth date is less than two years. As an illustration of the role of the features we present a sample from the iWiW data in Fig.~\ref{fig:egomy}. An egocentric network, i.e., the induced subgraph of an ego and her alters are shown, demonstrating the feature overlap. The traits of the four features of the ego are shown with different colored circle sectors. If a trait of an alter matches that of the ego, it is plotted with the same colored circle sector, otherwise if the alter's trait does not match that of the ego (e.g., the alter lives in a different city), then that sector is colored gray. We observe that there are hardly any alters who do not have any matching feature with the ego and they are most likely to be kins of the ego. It is worth noting that alters in different communities mainly have different sets of features in common with the ego.

Figure~\ref{fig:egomy}(a) was plotted using the Fruchterman-Reingold force-directed algorithm~\cite{fruchterman1991graph}, which pulls the connected nodes together while keeping the nodes at distance. Thus, communities of strongly connected nodes become visible. We identify communities in the egocentric network using the Infomap method~\cite{rosvall2009map}. In Fig.~\ref{fig:egomy}(b) it can be observed that the nodes in the largest community have surprisingly large number of matching features.

\section{Measures}

In order to quantify the similarity of an ego to her alters in terms of their features, we first define a \textit{feature overlap} function $\Delta(\sigma_i^f,\sigma_j^f)$. This function has a value of $1$ for matching traits of the $f$th feature between two connected users $i$ and $j$, and a value of $0$ for mismatching traits. If the feature is denoted by a categorical variable such as gender or city (marked with `cat' in Table~\ref{table:1}), the function $\Delta$ results in $1$ only when traits of two users match exactly (it is then equivalent to the Kronecker delta). If the feature is a numeric variable of traits (marked with `num' in Table~\ref{table:1}), then we allow the following difference in the matching feature: 
$\pm2$ years difference for age and $\pm1 \mathrm{kg/m}^2$ precision for BMI.

We define the \textit{link feature overlap} between users $i$ and $j$ by the mean of the feature overlap between them only using the features that exist for both users:
\begin{equation}
    o_{ij}\equiv \frac{1}{|F_{ij}|}\sum_{f \in F_{ij}} \Delta(\sigma_i^f,\sigma_j^f),
    \label{eq:overlap_link}
\end{equation}
where $F_{ij}$ is the set of features available for both users $i$ and $j$, and $|F_{ij}|$ is the cardinality of the set, in other words it is the number of features both $i$ and $j$ have. Only for the mobile phone data it may happen that $F_{ij}$ is an empty set, in which case these users are removed from the analysis.

We characterize the similarity of an ego, say $i$, to a subset $\Lambda$ of her alters in terms of the \textit{subset feature overlap} which is an average of $o_{ij}$ in Eq.~\eqref{eq:overlap_link} for $j\in \Lambda$:
\begin{equation}
o_i(\Lambda)\equiv\frac{1}{|\Lambda|}\sum_{j\in \Lambda} o_{ij}.
\label{eq:overlap_set}
\end{equation}
Here we consider two cases for the subset $\Lambda$: (i) $\Lambda= N_i$, i.e., all alters of the ego, and (ii) $\Lambda= C_i^r$, i.e., alters in the $r$th egocentric community of the ego $i$. In this study, we use the two-level Infomap method~\cite{rosvall2009map} to detect egocentric communities on the induced subgraph of the ego and her alters. Then we focus on an average value of subset feature overlaps in Eq.~\eqref{eq:overlap_set} over all communities of size $s$ for all egos:
\begin{align}
    \langle o\rangle(s)\equiv \langle o_i(C_i^r) \rangle_{\{i,r |s=|C_i^r|\}},
    \label{eq:overlap_commsize}
\end{align}
which we call a \emph{community feature overlap}. Similarly, we define an \emph{egocentric feature overlap}, by taking an average of subset feature overlaps in Eq.~\eqref{eq:overlap_set} for all egos having the same degree $k$ as follows:
\begin{align}
    \langle o\rangle(k)\equiv \langle o_i(N_i) \rangle_{\{i|k=|N_i|\}}.
    \label{eq:overlap_degree}
\end{align}

\begin{figure}[t]
    \centering
    \includegraphics[width=\columnwidth]{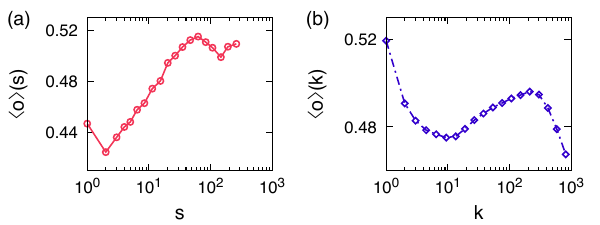}
    \caption{Empirical results of the iWiW data. (a) The community feature overlap as a function of community size $\langle o\rangle(s)$ in Eq.~\eqref{eq:overlap_commsize} and (b) the egocentric feature overlap as a function of the ego's degree $\langle o\rangle(k)$ in Eq.~\eqref{eq:overlap_degree}. Standard errors are not shown as they are smaller than the symbol size.
    }
    \label{fig:iwiwoverlaps}
\end{figure}

Throughout our analysis the degree of a node will be of crucial importance. It was already shown in Ref.~\cite{Torok2016What} that the degree of a node in an OSN is not determined by its number of friends in real life but rather, it might be the result of the inherent sampling of the real social network by the OSN. Moreover during the lifetime of the OSN the degree of a user tends to gradually increase. Therefore, we assume that the friends of a user in the datasets analyzed are a subset of its real life counterpart. But this subset is selected in a specific order characterized by the importance of the nodes to our social life.

Regarding the appearance order of alters in egocentric networks, we first chronologically sort out alters in each egocentric community for all egos; if an alter $j$ in an egocentric community $C_i^r$ of size $s$ appeared as the $n$th, the $j$'s appearance order is denoted by $a_i^r(j)=n$ for $n=1,\ldots,s$. Then one can define an \emph{appearance-order feature overlap} between egos and their $n$th appeared friends in all egocentric communities of size $s$:
\begin{align}
    \langle o\rangle_s(n)\equiv \langle o_{ij} \rangle_{\{i,j|s=|C_i^r|\ \textrm{and}\ a_i^r(j)=n\}}.
    \label{eq:overlap_order}
\end{align}

\section{Empirical results}

In this Section we report empirical results only for the iWiW data, while the other two datasets are discussed later. 

One of the important features of egocentric networks is their community structure, as it forms the basis of our social life and represents various facets of our lives. In our sample egocentric network in Fig.~\ref{fig:egomy} we can see that some communities have more overlap with the ego than others and here we would like to quantify this observation. In the case of iWiW data we could identify the users who had reliable representation of their real social network~\cite{Torok2016What}. By choosing users with at least 7 years of activity or with the number of alters between 150--250 we perform the egocentric community detection to obtain the community feature overlap as a function of the community size $s$, i.e., $\langle o\rangle(s)$ in Eq.~\eqref{eq:overlap_commsize}. In Fig.~\ref{fig:iwiwoverlaps}(a) the result shows an increasing trend up to $s \simeq 50$, after which the curve slightly decreases.

In Fig.~\ref{fig:iwiwoverlaps}(b) we show the egocentric feature overlap as a function of the ego's degree $k$, i.e., $\langle o\rangle(k)$ in Eq.~\eqref{eq:overlap_degree}. We find a local minimum at $k\simeq 10$ and a local maximum around at $k\simeq 200$, coinciding with the average degree of the network. The feature overlap around $k\simeq 200$ has the value of 
$\simeq 0.5$ that is comparable to the average feature overlap found in some large egocentric communities. 

We find both findings counter-intuitive. One would expect smaller groups to be potentially more uniform than large ones just for statistical reasons, as it is easier to find a handful of friends with social features similar to the ego than a few dozen matching ones. Moreover, naive thinking suggests that on average feature similarity correlates with emotional closeness. Thus, if we reach our friends in the OSN in the order of emotional closeness, then no peak other than the one at $k=1$ should be visible in the curves of $\langle o\rangle(k)$.

\begin{figure}
    \centering
    \includegraphics[width=0.75\columnwidth]{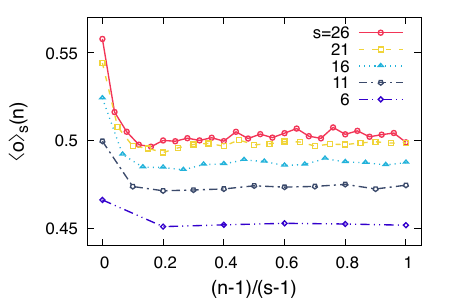}
    \caption{Empirical results of the iWiW data. We show the appearance-order feature overlap $\langle o\rangle_s(n)$ in Eq.~\eqref{eq:overlap_order} for several values of $s$. Here the order $n$ is scaled with the community size $s$ for clearer visualization.}
    \label{fig:comm_o_ranked}
\end{figure}

Next we look inside the communities and check how the overlap is distributed for alters appeared at different times. For this, we observe the evolution of egocentric networks in terms of the number of alters added; precisely, we measure the appearance-order feature overlap $\langle o\rangle_s(n)$ in Eq.~\eqref{eq:overlap_order}. In Fig.~\ref{fig:comm_o_ranked} we depict the feature overlaps of alters in the communities in the order of their appearance that are averaged over $25$--$160$ thousand communities. The appearance order has been scaled by the size of the community for clearer visualization. The curves for the larger communities are above the smaller ones as expected from Fig.~\ref{fig:iwiwoverlaps}(a). However, there is an interesting feature in the curves namely that the first few alters have significantly higher overlap than the rest having quite similar overlap values. It seems that in the egocentric communities only a few alters are really important for the ego. They are the ones with whom we connect to if we want to reach a community. Once we have reached a community the rest of alters are connected in a seemingly arbitrary order, i.e., irrespective of their feature overlaps with the ego.

Let us reiterate our two important findings: (i) the feature overlap of alters in egocentric communities increases with the community size and (ii) the feature overlap of alters appeared in the community is unevenly distributed such that on average only the first one or two members are really close to us while the rest have less but almost uniform importance. These are our basic observations, which we will be in the core of our model to explain other results.

\begin{figure*}[t!]
    \centering
    \includegraphics[width=0.75\textwidth]{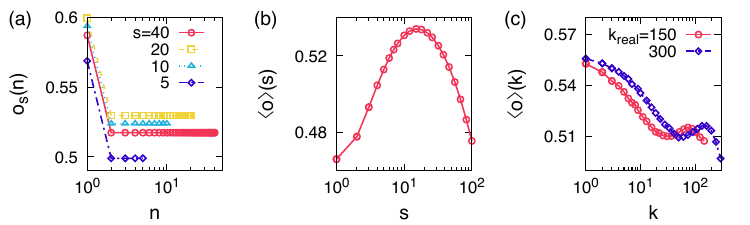}
    \caption{Model assumption and results: (a) Appearance-order feature overlap curves assumed for the model in Eq.~\eqref{Eq:osn_model} for several community sizes $s$. (b) The resultant curve of $\langle o\rangle (s)$ given by Eq.~\eqref{eq:os_model}. (c) Simulation results of the egocentric feature overlap as a function of the ego's degree for the cases with $k_{\rm real}=150,300$.}
    \label{fig:simplemodelres}
\end{figure*}

\section{Model}

We devise a simple model for testing the empirical findings in the previous Section. It is assumed that the distribution of community sizes follows a power law with exponent $-1.5$. This exponent value is based on the analysis of iWiW data. Precisely, we assume that
\begin{align}
    P(s)= A s^{-1.5}\ \textrm{for}\ 2\leq s\leq 100,
    \label{eq:Ps}
\end{align}
with a normalization constant $A=1/(\sum_{s=2}^{100} s^{-1.5})$. Let us consider an ego whose real-world degree is $k_{\rm real}$. Then we draw a number of community sizes from $P(s)$ in Eq.~\eqref{eq:Ps} such that the sum of those sizes equals to $k_{\rm real}$. Let the ego join the OSN and start to make connections to alters from those communities one by one until the ego's degree $k$ in the OSN reaches $k_{\rm real}$. Then, users of the OSN whose degree is $k$ can be seen as egos who have sampled $k$ alters from the real-world alters. In our model, we considers the cases with $k_{\rm real}=150,300$.

Next, we model the appearance-order feature overlap for alters in communities of the ego. For a community of size $s$, the ego's $n$th alter in the community is assumed to have the feature overlap with the ego as follows:
\begin{align}
\label{Eq:osn_model}
o_s(n)=\left(1-\frac{1}{s+2}\right)\left(\frac{60}{s^{0.7}+100}\right)+0.07\delta_{n,1} 
\end{align}
for $n=1,\ldots,s$, where $\delta$ is a Kronecker delta. The first term on the right hand side of Eq.~\eqref{Eq:osn_model} is an increasing and then decreasing function of community size $s$ for the same $n$. It reflects the observation in Fig.~\ref{fig:iwiwoverlaps}(a). The second term on the right hand side of Eq.~\eqref{Eq:osn_model} is to demonstrate the observation that the first appeared alter in each community tends to have the larger feature overlap with the ego than the rest, see Fig.~\ref{fig:comm_o_ranked}. Here the functional form and constant values are not fitted from the data, but chosen for the demonstration of the model. Some typical functions are plotted in Fig.~\ref{fig:simplemodelres}(a).

Using the appearance-order feature overlap of the model in Eq.~\eqref{Eq:osn_model}, we calculate the community feature overlap as
\begin{align}
    \langle o\rangle(s)=\frac{1}{s}\sum_{n=1}^s o_s(n).
    \label{eq:os_model}
\end{align}
The numerical result of Eq.~\eqref{eq:os_model} is shown in Fig.~\ref{fig:simplemodelres}(b), which is qualitatively similar to the empirical result in Fig.~\ref{fig:iwiwoverlaps}(a).

Now let us grow the egocentric network using both generated communities and the appearance-order feature overlap curves for those communities. We start with an isolated ego.
\begin{enumerate}
    \item A community, say $r$, is randomly chosen from the set of communities.
    \item Among remaining individuals in the community $r$, we take the individual having the largest feature overlap with the ego, which becomes the ego's alter.
    \item That alter is removed from the community $r$. If the community $r$ becomes empty, it is removed from the set of communities.
    \item The above procedure (1--3) is repeated until the ego's degree reaches $k_{\rm real}$.
\end{enumerate}
Using this procedure we generate 5,000 egocentric networks to take the average of the feature overlaps of the $k$th alters of all egos. Then we get $\langle o\rangle(k)$, which is shown in Fig.~\ref{fig:simplemodelres}(c). The curve $\langle o\rangle(k)$ starts with a large value; the first alters have higher feature overlaps with the ego because each of them is the first appeared alter in each community. Then the curve decreases as the rest of alters in communities who have lower values of the feature overlap start to be added to the egocentric network. Once alters in smaller communities with smaller feature overlap are all added, then only alters in larger communities are left. Thus for larger $k$ we find the increasing tendency of the $\langle o\rangle(k)$ curve. Finally, when the nonsocial alters in very large communities are added to the egocentric network, $\langle o\rangle(k)$ decreases again. This is why we have both local minimum and local maximum in the $\langle o\rangle(k)$ curve.

\begin{figure}[t!]
    \centering
    \includegraphics[width=\columnwidth]{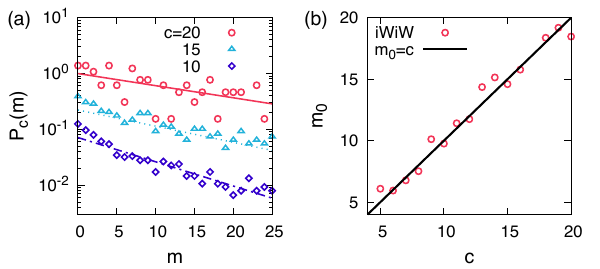}
    \caption{Empirical results of the iWiW data. (a) We show probability distribution functions $P_c(m)$ of the chronological order $m$ of the alter who appeared in the egocentric network for the first time among members of each community for egos having $c$ communities. Theoretical expectations of $P_c(m)\propto e^{-m/c}$ are shown in lines. Empirical results and expected curves for $c=15$ and $20$ are shifted for clearer presentation. (b) Estimated values of the scale parameter $m_0$ in Eq.~\eqref{eq:Pcm_data} as a function of $c$. For comparison, the expectation $m_0=c$ by Eq.~\eqref{eq:Pcm} is plotted in black line. 
    }
    \label{fig:positionexp}
\end{figure}

The model introduced here reproduces qualitatively the empirical findings, but the assumption that the communities are chosen uniformly at random needs to be further tested. Consider an ego whose egocentric network has $c$ communities. We can track the chronological order of alters appeared in the egocentric network and which community each alter belongs to. That is, we know the chronological order $m$ of the alter who appeared in the egocentric network for the first time among members of each community, leading to exactly $c$ values of $m$ for each ego. By collecting such $m$s for all egos whose egocentric networks have $c$ communities, we get the probability distribution function $P_c(m)$. If our assumption that communities are randomly chosen is correct, we expect to have
\begin{align}
    P_c(m)=\left(1-\frac{1}{c}\right)^{m-1}\frac{1}{c}\approx \frac{1}{c}e^{-m/c},
    \label{eq:Pcm}
\end{align}
where the approximation has been made by assuming that $m,c\gg 1$.

The distribution $P_c(m)$ can be directly measured from the iWiW data as the chronological orders of alters of given egos and communities alters belong to are available. As shown in Fig.~\ref{fig:positionexp}(a) the distribution $P_c(m)$ turns out to be exponential for the range of $m\leq 25$ for several values of $c$, implying that our assumption is reasonable. By assuming that 
\begin{align}
    P_c(m)\propto e^{-m/m_0}
    \label{eq:Pcm_data}
\end{align}
with a scale parameter $m_0$, we estimate the value of $m_0$ for each value of $c$. It is expected that $m_0=c$ by Eq.~\eqref{eq:Pcm}. The results are depicted in Fig.~\ref{fig:positionexp}(b), showing that our expectation $m_0=c$ is indeed the case. Note that communities have finite number of members and once all members in a community are added to the egocentric network, then that community cannot be selected any more. Thus $P_c(m)$ becomes no more exponential. This is why we fit the empirical distribution for the range of $m\leq 25$ to obtain convincing exponential functions. 

We can summarize our major observations: The average egocentric community overlap increases with increasing community size, as we add friends to an egocentric community, such that first someone close to us is reached then the rest is added seemingly random order, irrespective of the size of the communities. 

\section{Discussion}

We have shown that on average, in the egocentric communities the ego has larger feature overlap with  alters in the larger communities. At first, this result might be surprising since it is always more difficult to find more people who are similar to us. Thus populating a large community with very similar friends would seem more difficult than a smaller one. However, there are arguments for this observations. Social links fade in time and they need regular interactions to remain active~\cite{Murase2015Modeling}. Thus in order to keep up the many links of a large community more activity is needed, which in turn requires more social similarity~\cite{Murase2019Structural}. As we reach out to the communities the first one or two close friends are connected, after which the rest occurs at random. Using the empirical data from an online social network we have shown that as the ego populates its egocentric network with her alters she chooses the communities randomly. This results in an egocentric feature overlap, which has a local minimum at around $k=12$--$15$ a number very close to the size of the Dunbar's circle for close friends. If we identify the local minimum of the overlap degree curve with the second Dunbar's circle it would mean that our good friends are not only the emotionally closest friends but they are the backbone of our social life, meaning that they connect us to our social activities. This is meaningful as usually all aspects of our social life are important to us and these are the people who connect us to them. The actual minimum of the egocentric feature overlap curve is at $k=13$ which coincides with the average number of egocentric communities a user has.

For large degrees the empirical data shows a decreasing trend in the overlap after $\sim$$200$ due to the increasing fraction of nonsocial contacts. This is close to the Dunbar's number. Our data indicates that indeed after around $\sim$$200$ contacts the nature of the ties change.

\section{Other datasets}

\subsection{Pokec data}

Pokec is an existing Slovakian social network site. The dataset~\cite{Takac2012Data} contains profile information for some 1.6 million users for a country 
with 5.5 million inhabitants. It also includes connection data which is not timestamped. The profile contains many information fields. We used the ones for which at least 30\% of the users gave meaningful answer. Some fields such as alcohol, sex, smoking related habits were free strings. There were 3--5 answers with identical strings which were enumerated in our work. For examples, we considered `abstinent' as it is, `pijem prilezitostne' as ``I drink occasionally", `pijem pravidelne' as ``I drink regularly", `pijem' as ``I drink". The BMI was calculated from the given height and mass values and only reasonable values between 10 and 50 were kept. The location was given by counties. The two major cities, Bratislava and KoÅ¡ice, are divided into sub regions, implying that these were not aggregated into one city.

\begin{figure}[t!]
    \centering
    \includegraphics[width=\columnwidth]{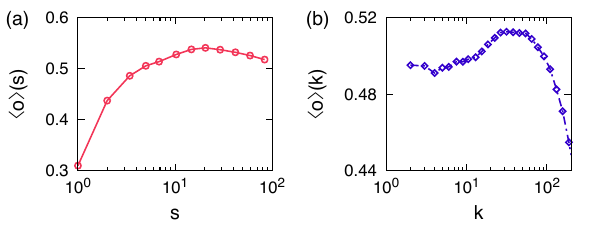}
    \caption{Empirical results of the Pokec data: (a) the community feature overlap as a function of community size $\langle o\rangle(s)$ in Eq.~\eqref{eq:overlap_commsize} and (b) the egocentric feature overlap as a function of the ego's degree $\langle o\rangle(k)$ in Eq.~\eqref{eq:overlap_degree}.}
    \label{fig:pokec}
\end{figure}

The community feature overlap is shown in Fig.~\ref{fig:pokec}, showing that the curve increases up to $s\simeq20$, which is comparable to the behavior of the iWiW data which has three times more users and average degree than Pokec. The egocentric feature overlap in Fig.~\ref{fig:pokec}(b) shows a maximum at around $k\simeq 30$--$50$. The initial peak is only marginal.

\subsection{Call Detail Records}

In addition to the OSN datasets we analyze the mobile phone dataset having 5,716,084 subscribers from a European country in the year 2007 whose demographic information is displayed in Table~\ref{table:1}. The users were anonymized by the service provider with unique identifiers such that the privacy of the individuals are protected and cannot be traced back to them. The call detail records (CDRs) include all the incoming and outgoing calls made by an individual during the study period. It also includes the age and gender of the caller and the callee which has been considered as the main features for feature overlap calculation in the present paper. We studied approximately 600,000 egocentric networks sampled randomly from the data and we were left with 123,533 egocentric networks with the egos and her alters having at least one known feature.

\begin{figure}[t!]
    \centering
    \includegraphics[width=\columnwidth]{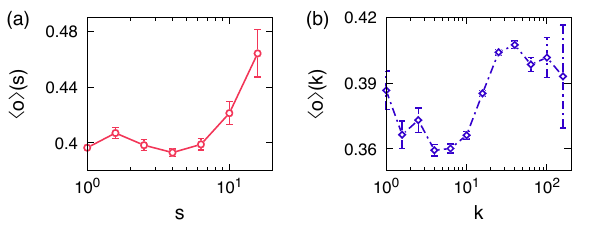}
    \caption{Empirical results of the CDRs: (a) the community feature overlap as a function of community size $\langle o\rangle(s)$ in Eq.~\eqref{eq:overlap_commsize} and (b) the egocentric feature overlap as a function of the ego's degree $\langle o\rangle(k)$ in Eq.~\eqref{eq:overlap_degree}. The error bars represent standard errors.}
    \label{fig:mob_overlap}
\end{figure}

In Fig.~\ref{fig:mob_overlap}(a) we show the community feature overlap as a function of community size. As shown in the figure, the range of community size is limited due to the nature of CDRs analyzed here. In our work the link between two users is considered meaningful only when they have called each other at least once during the period of study. Thus the average degree itself is quite small, leading to relatively small egocentric community sizes, in contrast to iWiW and Pokec datasets. We also note that since it requires an extra effort to make calls and talk using mobile phones we believe that the network from CDRs consists of more meaningful relationships, for example, important friends and family members, who are called frequently and on a regular basis. Moreover, the study period was considered for the year 2007 when mobile phones were still being used as a major form of communication channel and so this network represents fewer but more closer relationships. Even so, we find a similar tendency of the community feature overlap to the OSN datasets, namely, the community feature overlap increases from a community size 4 onwards. Figure~\ref{fig:mob_overlap}(b) shows the egocentric feature overlap. Here we considered all those neighbors whose at least one feature is known. We find the same patterns as the iWiW and Pokec datasets. 

\section{Summary}

In summary we have shown by analyzing three different empirical data sources that the larger the egocentric community the more similar its members are to the ego. As a user adds its friends to its egocentric network on an OSN the communities appear essentially in a random manner. We have also shown that in all egocentric communities we first connect to one or two really close friends while the rest on average has comparable emotional closeness. Finally we tested our empirical findings by devising a simple model that is able to reproduce all the mentioned empirical findings.

\begin{acknowledgments}
C.R. and K.K. acknowledge support from EU HORIZON 2020 INFRAIA-1-2014-2015 program project (SoBigData) No.~654024 and INFRAIA-2019-1 (SoBigData++) No.~871042. 
H.-H.J. acknowledges financial support by the National Research Foundation of Korea (NRF) grant funded by the Korea government (MSIT) (No.~2022R1A2C1007358).
J.K. acknowledges support from EU HORIZON 2020 INFRAIA-2019-1 (SoBigData++) No.~871042 and CHIST-ERA-19-XAI010 SAI projects, FWF (grant No.~I 5205).
J.T. thanks the support of NKFIH Hungarian Research Fund grant 134199 and of the NKFIH Fund TKP2021 BME-NVA-02, carried out at the Budapest University of Technology and Economics.
\end{acknowledgments}

\end{document}